\documentclass[12pt]{article}

  \usepackage{graphicx}
  \usepackage{epsfig}
  \usepackage{amssymb}
  \usepackage{subfigure}
  \usepackage{amsmath} 

\evensidemargin - .5truecm
\allowdisplaybreaks[1]

\def \lsim
{\raisebox{-3pt}{$\>\stackrel{<}{\scriptstyle\sim}\>$}}
\def \gsim
{\raisebox{-3pt}{$\>\stackrel{>}{\scriptstyle\sim}\>$}}

\begin{document}

\newcommand{\CC}{{\mathbb C}}
\newcommand{\RR}{{\mathbb R}}
\newcommand{\ZZ}{{\mathbb Z}}
\newcommand{\QQ}{{\mathbb Q}}
\newcommand{\NN}{{\mathbb N}}
\newcommand{\beq}{\begin{equation}}
\newcommand{\eeq}{\end{equation}}
\newcommand{\beal}{\begin{align}}
\newcommand{\eeal}{\end{align}}
\newcommand{\nn}{\nonumber}
\newcommand{\bea}{\begin{eqnarray}}
\newcommand{\eea}{\end{eqnarray}}
\newcommand{\ba}{\begin{array}}
\newcommand{\ea}{\end{array}}
\newcommand{\bfig}{\begin{figure}}
\newcommand{\efig}{\end{figure}}
\newcommand{\bc}{\begin{center}}
\newcommand{\ec}{\end{center}}

\newenvironment{appendletterA}
{
  \typeout{ Starting Appendix \thesection }
  \setcounter{section}{0}
  \setcounter{equation}{0}
  \renewcommand{\theequation}{A\arabic{equation}}
 }{
  \typeout{Appendix done}
 }
\newenvironment{appendletterB}
 {
  \typeout{ Starting Appendix \thesection }
  \setcounter{equation}{0}
  \renewcommand{\theequation}{B\arabic{equation}}
 }{
  \typeout{Appendix done}
 }

%%%%%%%%%%%%%%%%%%%%%%%%%%%%%%%%%%%%%%%%%%%%%%%%%%%%%%%%%%%%%%%%%%%%%%%%
%
%
%\begin{fmffile}{HIGAGA}
%
%
%%%%%%%%%%%%%%%%%%%%%%%%%%%%%%%%%%%%%%%%%%%%%%%%%%%%%%%%%%%%%%%%%%%%%%%%

\begin{titlepage}
\nopagebreak
\renewcommand{\thefootnote}{\fnsymbol{footnote}}
\vskip 2cm
\begin{center}
\boldmath
{\Large\bf 
Renormalization in Winter Model
}
\unboldmath
\vskip 1.cm
{\large  U.G.~Aglietti\footnote{Email: Ugo.Aglietti@roma1.infn.it},~~~
         P.M.~Santini\footnote{Email: Paolo.Santini@roma1.infn.it}, }
\vskip .2cm
{\it Dipartimento di Fisica, Universit\`a di Roma ``La Sapienza''  and
\vskip 0.1truecm
INFN, Sezione di Roma, 00185 Rome, Italy} 
\end{center}
\vskip 0.7cm

\begin{abstract}

We show that metastable states in Winter model can be related to the
eigenstates of a particle in a box by means of renormalization and mixing.


\vskip 7truecm

\end{abstract}

\vfill
\end{titlepage}    

\setcounter{footnote}{0}

\section{Introduction}

The theory of unstable states in quantum mechanics 
\cite{Gamow,refbase,refbase2,hatano,chap5,scatteringth} 
has applications virtually in any branch of physics:
statistical and condensed matter physics \cite{anderson},
atomic and molecular physics \cite{chap4},
nuclear physics \cite{segre,RGtonuclei}, 
quantum field theory and particle physics \cite{mt}, 
and so on.
In this note we show that the particle states inside the cavity 
of the Winter model \cite{winter,flugge,chap7,chap9,mailaggr,primo} 
can be obtained from the 
states of a particle in a box by means of renormalization  and mixing.
Winter model describes the coupling of a cavity with the outside and
is given, after proper rescaling \cite{primo} (see next section), 
by the Hamiltonian
\beq
\label{eqmod}
\hat{H} = -\frac{\partial^2}{\partial x^2} + \frac{1}{\pi g} \delta(x-\pi)
\eeq
in the half-line $0\le x < \infty$, and with vanishing boundary condition at zero:
$\psi(x=0,t)=0$. The distribution $\delta(y)$ is the Dirac $\delta$-function. 
Eq.(\ref{eqmod}) describes a model with one parameter,
$g\in\RR$, the inverse of the area of the potential barrier in $x=\pi$ (up to
a factor $\pi$).
The metastable states are nothing but wave packets initially (i.e. for example at $t=0$)
concentrated inside the cavity, i.e. in the interval $0<x<\pi$. 
The time evolution of metastable states is controlled by wave propagation and imperfect 
multiple reflections on the right cavity wall, in $x=\pi$, 
leading to a leakage of the wave amplitude outside it.
For $0< g \ll 1$ (high barrier), there is weak coupling of the cavity with the outside
and resonant long-lived states come into play.
The idea is that, by means of them, we can describe the dynamics of the particles
initially inside the cavity {\it as if} the outside did not exist. 

Let us briefly discuss the motivations which led us to further investigate
Winter model.

\noindent
Historically, this model has been used for a semi-quantitative analysis
of $\alpha$-decay in heavy nuclei (see \cite{segre} and references therein), 
as the superposition of the nuclear and electrostatic potential can be roughly 
described by a bump function (a general potential with compact support can be 
approximated with a $\delta$-function for large wavelengths). 
Of course, there are clearly much more realistic 
models in this context, to be analyzed with numerical methods.

\noindent
Winter model is currently used in quantum chemistry to check metastable 
properties of more complex models describing some specific phenomenology 
\cite{chap7,chap9}, as a source for exact quantum-decay solutions 
\cite{mailaggr}, or as a testing ground for various resonant-state 
formalisms (Green functions, spectral decompositions, etc. \cite{refbase}).  
Its relevance lies in the fact that it is the simplest Hamiltonian system 
exhibiting metastable states, for which an almost exact analytic treatment 
is possible. On the contrary, more phenomenologically oriented models, 
involving for example a sequence of delta-like barriers or 
additional potentials modeling some media or detector interaction 
\cite{refbase}, have to be analyzed by means of specific numerical 
techniques. 
In this framework, Winter model provides a safe and simple check of 
results obtained for more complex models, being a limiting case. 

\noindent
There is also a link of Winter model to a big and active field of 
current research: Quantum Chromodynamics on the Lattice. 
The box eigenfunctions play the role in Winter model 
of the interpolating fields in quantum field theory. 
The latter are operators acting on the vacuum by exciting a multiplicity of 
different particle states, which one is usually interested to separate, 
as we explicitly can in Winter model by means of the inverse of
the mixing matrix $U^{-1}(g)$ (see later). 

\noindent
Finally, Winter model also has a pedagogical relevance, as it enables
a student in physics or chemistry to fully understand the decay of
states in quantum mechanics without the use of perturbation theory
(Fermi golden rule). Indeed, by using perturbative methods, one is
always faced with the problem of separating what is intrinsic to the 
system from what is instead just a consequence of the approximations.
On the theoretical side, the application of perturbation theory to
decay phenomena is not straightforward (while of course correct), 
because an arbitrarily small perturbation produces a drastic change 
in the spectrum of the theory: the discrete spectrum completely disappears.
In the same spirit, resonance properties of Winter model for small 
$|g|$ were discussed in the book of applications of quantum mechanics 
by S.$\,\,$Flugge \cite{flugge}.

Our results involve a second-order computation in $g$ extending the $\mathcal{O}(g)$ 
results in \cite{primo}, in which we repeated the original Winter's computation, 
finding additional contributions in the time evolution of unstable states, 
which were absent in \cite{winter}. 
These new terms have a small strength $\mathcal{O}(g) \ll 1$ compared
to the old ones, but decay generally slower in time, with the smallest decay width.
These contributions therefore dominate the evolution of all unstable states except
the lowest one at large times and cannot be neglected.
In this note we show that such contributions can be in principle ``rotated away''
by means of a linear transformation $U(g)$ in the infinite-dimensional vector space
of the resonances. 

Since Winter model is, as discussed above, a kind of "attractor'' of 
many quantum-decay models, it is clear that the occurrence of nondiagonal terms 
has to be a general phenomenon in metastable systems. 
The implications of the mixing terms for the checks discussed above are, 
as far as we know, still to be investigated.  
Let us stress that it would have been difficult to imagine the existence of such 
mixing terms in complex models without an exhaustive analytic study of Winter model. 

To show the relevance of "renormalizing'' Winter model, let us briefly discuss the
importance of renormalization in understanding the dynamics of many physical theories. 
It is common practise in physics to compare a given physical system with a simplified
one in which some interactions are omitted. These interactions can be related to control 
parameters in experiments involving, for example, external electric or magnetic fields,
or can be treated theoretically as variable quantities.
The idea of renormalization is that switching on an interaction in a physical
system has, as one of the main dynamical effects, that of modifying the 
parameters of the starting, 
non-interacting, system; once renormalization has been made, the residual effects of the 
interaction are substantially weaker than before renormalization. 
When the coupling setting the strength of the interaction gets too large, 
there is usually little connection between the free and the interacting system
and renormalization often looses its meaning.

In condensed matter physics, renormalization is related to the so called
``adiabatic continuity principle'' \cite{anderson}: by adiabatically (i.e. slowly)
turning on an interaction, the free states of the system go $1-1$ onto the
interacting states by means of a flow of the parameters such as masses, couplings,
etc. A typical example is the normal Fermi liquid, i.e. a system with a repulsive 
interaction among electrons in the Fermi sphere
\footnote{
See footnote \ref{nfl}.  
}.
In quantum field theory \cite{hint,lepage,ms,zinnjustin}, 
the relation between free parameters and interacting ones (masses, couplings
and field normalizations)
is often singular because of the lack of intrinsic energy scales cutting off
the quantum fluctuations at large energies: that leads to the well-known 
ultraviolet infinities. One also encounters renormalization in 
non-relativistic quantum mechanics with $\delta$-functions potentials 
\cite{jackiw}.

It is remarkable that solving the non-relativistic Schrodinger equation
for nuclei (which are strongly-interacting many-body systems), 
in order to obtain the low-energy excitations and the scattering cross
sections, can be greatly simplified by implementing renormalization-group
ideas, as recently discovered \cite{RGtonuclei}.
One finds the phenomenon of generation of many-body operators by RG flow, 
the problem of the stability under change of the ultraviolet cutoff, etc.,
which are typical of perturbative quantum field theory computations,
in a completely different framework.

In classical physics, renormalization is usually implemented by the
method of ``multiple scales'' \cite{landau1,ingegnere}. In the case of a
free anharmonic oscillator, for example, renormalization amounts to the
``absorption'' of secular terms into a shift of the harmonic 
frequency. These terms are formally resonances produced by
forcing terms occurring in the perturbative expansion, 
are incompatible with energy conservation and spoil the 
convergence of the perturbative expansion at large times.  
After renormalization, such strong-coupling effects completely
disappear; only a small coupling between the
harmonics is left and a uniform approximation in time is obtained.

Let us remark that the adiabatic continuity principle is subjected
to relevant violations. Let us quote, for example, the cases of
the energy gap in the Bardeen-Cooper-Schriffer (BCS) theory of 
classical superconductors \cite{bcs} or the mass gap in massless 
Quantum-Chromo-Dynamics (QCD)  \cite{bochicchio}.
These phenomena are typically characterized by functions which have
an essential singularity when the interaction coupling $g$ goes to zero,
of the form $e^{1/g}$ for $g<0$, 
making non smooth the connection between the interacting system and the 
related free one.  
In these cases, the relation between the free system and the interacting one is 
highly non trivial and the residual interaction is of nonperturbative character.

Let us end the introduction by observing that, even though renormalization is 
implemented and interpreted in quite different ways in different contexts, 
it is an ubiquitous phenomenon in physics --- like the unstable states cited above ---
a thing which certainly could not be expected {\it a priori}.

\section{Winter Model}

The Hamiltonian operator of Winter model reads:
\beq
\label{WinterH}
\hat{H} \, = \, - \, \frac{ \hbar^2 }{2 m} \frac{\partial^2}{\partial x^2} 
\, + \, \lambda \, \delta(x - L) \, ,
\eeq
where $m$ is the particle mass, $\lambda$ is a coupling constant, $\delta$ is the
Dirac $\delta$-function and $x=L>0$ is the support of the potential.
Formulae can be simplified by going to a proper adimensional coordinate via
\beq
x = \frac{L} {\pi} x'
\eeq
and rescaling the Hamiltonian as:
\beq
\hat{H} =  \frac{ \hbar^2 \pi^2 }{ 2 m L^2 } \hat{H}'.
\eeq
The new (adimensional) Hamiltonian then takes the form in which it
appeared in the introduction:
\beq
\label{rescaled}
\hat{H'} \, = \, - \, \frac{\partial^2}{\partial x'^{\, 2} } \, + \, \frac{1}{\pi g}  \, \delta(x' - \pi),
\eeq
and contains the single real parameter 
\beq
g \, = \, \frac{ \hbar^2  }{ 2 m \lambda L } .
\eeq
The time-dependent Schrodinger equation
\beq
i \hbar \frac{\partial \psi }{\partial t }  = \hat{H} \psi  
\eeq
now reads
\beq
i \frac{\partial \psi }{\partial t' }  = \hat{H'} \psi ,
\eeq
where
\beq
t' \equiv  \frac{ \hbar \pi^2 }{ 2 m L^2 } t .
\eeq
Let us omit primes from now on for simplicity's sake.
It is possible to rescale Winter Hamiltonian (\ref{WinterH}) in slightly
different ways, as made for example by Winter itself, but the main point 
is that we deal in any case with a one-parameter model. 

\subsection{The spectrum}

For a positive coupling, $g>0$, i.e. for a repulsive potential,
the Hamiltonian of Winter model in eq.(\ref{eqmod}) has a continuum spectrum
only, with eigenfunctions of the form
\bea
\label{nondiviso}
\psi(x;k,g) &\propto& 
\Big[ - \frac{i}{2} \, \exp(i k x)   \, + \, \frac{i}{2} \, \exp(- i k x) \Big] \, 
\theta(\pi - x)
\, + \,  
\nonumber\\
&+& 
\Big[
a(k,g) \, \exp(i k x) \, + \, b(k,g) \, \exp(- i k x)
\Big]
\, \theta(x -  \pi)  \, ,
\eea
and eigenvalues 
\beq
\epsilon(k) = k^2 . 
\eeq
The step function $\theta(x)=1$ for $x>0$ and $0$ otherwise
and the coefficients $a(k,g)$ and $b(k,g)$ have the following
expressions:
\bea
a(k,g) &=& - \, \frac{i}{2} \, + \, \frac{1}{4\pi g k} \left[ \exp( - 2 i \pi k ) - 1 \right] \, ;
\\
b(k,g) &=& + \, \frac{i}{2} \, + \, \frac{1}{4 \pi g k} \left[ \exp( + i 2 \pi k ) - 1 \right] \, .
\eea
These coefficients have the following two symmetries (which will
be relevant in the discussion of the spectrum as well as of the time
evolution): 
\bea
\label{sim1}
a(-k,g) &=& - b(k,g) ,
\\
\label{sim2}
a(k,g)^* &=& b(k^*,g^*) ,
\eea
where the star $(^*)$ denotes complex conjugation.
In general, $k$ is a real quantum number but, because of eq.(\ref{sim1}),
the eigenfunctions are odd functions of $k$ so one can assume $k>0$,
implying there is no energy degeneracy (trivial $S$-matrix).
By normalizing the eigenfunctions as:
\beq
\label{continuo}
\int_0^\infty \psi^*(x;k',g) \, \psi(x;k,g) \, dx \, = \, \delta( k - k' ) \, ,
\eeq
where $\delta(q)$ is the Dirac $\delta$-function, the normalization factor reads:
\beq
{\cal N}(k,g) \, = \, \frac{1}{ \left[ 2\pi a(k,g) b(k,g) \right]^{1/2}  } . 
\eeq
The final expression for the eigenfunctions therefore can be written as:
\bea
\label{diviso}
\psi(x;k,g) &=&
\frac{1}{\sqrt{2\pi}}
\left[
\left( 
- \frac{i \, \exp(i k x)}{2 \left[ a(k,g) b(k,g) \right]^{1/2} }   \, + \, \frac{i \, \exp(- i k x)}{2  \left[ a(k,g) b(k,g) \right]^{1/2} } 
\right) 
\, \theta(\pi - x)
\, + \, \right.
\nonumber\\
&& \, \, \, \, + 
\left.
\left(
\left[ \frac{ a(k,g)}{b(k,g)} \right]^{1/2} \, \exp(i k x) \, + \, \left[ \frac{b(k,g)}{a(k,g)} \right]^{1/2} \, \exp(- i k x)
\right)
\, \theta(x -  \pi)  
\right]. \, 
\eea
Note that, because of continuum normalization, the amplitude of the eigenfunctions 
outside the wall is always $\mathcal{O}(1)$, no matter which values
are chosen for $k$ and $g$, while inside the cavity the amplitude has a 
non-trivial dependence on $k$ and $g$.
For $|g| \ll 1$, the amplitude of $\psi(x;k,g)$ inside the cavity shows marked
peaks for $k \approx n - g n$, where $1 \le n \ll 1/|g|$ is an integer, because
\beq
|a(n - g n,g)| \, = \, |b(n - g n,g)| \, = \, \frac{|g|}{2} \sqrt{ \pi^2 n^2 + 1} 
+ \mathcal{O}\left(g^2\right)
\, \approx \,  \frac{\pi}{2} |g| n \, \ll \, 1 .
\eeq
As is usually the case, peaks become less marked for increasing $n$.

\section{Temporal evolution of unstable states}

The eigenfunctions of a particle in a box of length $L=\pi$ with Hamiltonian
\beq
\hat{H}_0 = -\frac{\partial^2}{\partial x^2}
\eeq
are given, as well known, by 
\beq
\label{freeeg}
\psi^{(l)}_0(x,t) = \sqrt{\frac{2}{\pi}} \sin (l x) e^{-i l^2 t} ,
\eeq
where $l=1,2,3,\cdots$ is a positive integer and $0\le x \le \pi$.
We study the time evolution of wavefunctions $\psi^{(l)}(x,t;g)$ which coincide
at $t=0$ with the free eigenfunctions in eq.~(\ref{freeeg}) in the interval $x\in [0,\pi]$
(the cavity) and vanish outside it:
\beq
\label{initial}
\psi^{(l)}(x,0) \, = \, 
\left\{
\begin{array}{cc}
\sqrt{ 2/\pi } \, \sin \left( l \, x \right) & {\rm for} ~ 0 \le x \le \pi ; ~~
\\
0 &{\rm for} ~ \pi < x < \infty \, .
\end{array}
\right.
\eeq
The initial conditions above make the limit $g \to 0$ easy, because the 
wavefunctions $\psi^{(l)}(x,t;g)$ become eigenfunctions of Winter Hamiltonian
in that limit. 
For $g\ne 0$, however, we will see in the next section that there are more
natural initial conditions to consider.

\noindent
The spectral representation in eigenfunctions of the wavefunction of the unstable state
at time $t$ has the explicit expression:
\beq
\label{psi}
~~~~~~~
\psi^{(l)}(x,t;g) \, = \, 
\left(\frac{2}{\pi}\right)^{3/2} \int\limits_{0}^{\infty} p^{(l)}(k; x, g) \, e^{-ik^2t} \, dk , 
~~~~~~~\, \,  0 \, \le \, x \, \le \, \pi \, , ~~~\, \,  g \, > \, 0 \, ,
\eeq 
where
\beq
p^{(l)}(k;x,g) \, = \, (-1)^l l \frac{\sin (k \pi)}{ k^2 - l^2 } 
\, \frac{\sin (k x)}{ 4 a(k,g) b(k,g) } \, .
\eeq
The integral on the r.h.s. of eq.(\ref{psi}) can be exactly evaluated with
numerical methods for $t$ not too large, because high-frequency oscillations
occur in the factor $e^{-ik^2t}$ in the integrand for $t \to + \infty$.
In order to study the large-time behavior we therefore have to develop 
analytic techniques.

\subsection{Small-time Expansion}

For small times, $t\ll 1$, the wavefunction in eq.(\ref{psi}) exhibits a power
behavior \cite{chap4,winter,primo} that is not relevant to our discussion
and will not be treated further.

\subsection{Asymptotic Expansion for large times}

\begin{figure}[ht]
\begin{center}
\includegraphics[width=0.5\textwidth]{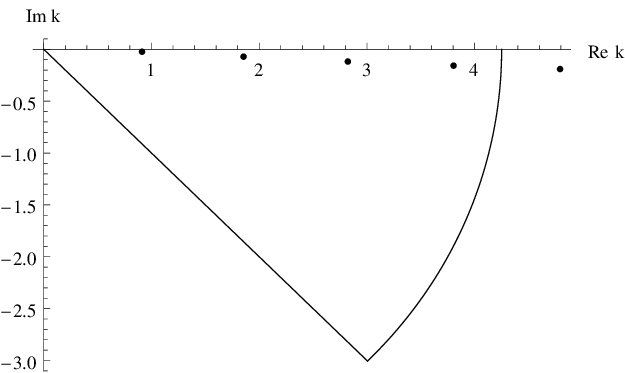}
\footnotesize\caption{
\label{figura1}
\it Rotation of the integration contour in the complex $k$-plane (see text) 
and (simple) zeroes of the function $b(k,g)$ for $g=0.1$ lying in the 
fourth quadrant.}
\end{center}
\end{figure}

To obtain explicit analytic formulae, we expand the integral for large $t$. 
The steepest descent method suggests to replace the integral on the r.h.s. of eq.~(\ref{psi}) 
by the integral over the steepest descent ray ($0,\infty e^{-i\pi/4}$), on which 
the fast oscillations of the integrand are absent (see fig.\ref{figura1}). 
Therefore the state $\psi^{(l)}(x,t;g)$ is decomposed in a natural way into the sum of two quite
different contributions:
\beq
\label{residue}
\psi^{(l)}(x,t;g) \, = \, \psi^{(l)}_{ex}(x,t;g) \, + \, \psi^{(l)}_{pw}(x,t;g) \, ,
\eeq
where
\bea
\psi^{(l)}_{ex}(x,t;g) &\equiv& - 2 \pi i\left(\frac{2}{\pi}\right)^{3/2}\sum_{n=1}^{\infty} {\rm Res} 
\left[ p^{(l)}(k; x,g) \, e^{- i k^2 t} , \, k^{(n)}(g) \right] \, ;
\\
\label{powercorr}
\psi^{(l)}_{pw}(x,t;g) &\equiv& 
e^{-i\pi/4}\left(\frac{2}{\pi}\right)^{3/2} \, \int\limits_{0}^{\infty} p^{(l)}\left( k \, e^{-i\pi/4}; x,g \right) 
\, e^{-k^2 t} \, dk \, .
\eea    
${\rm Res}\left[f(k); \, a \right]$ denotes the residue of the
(analytic) function $f$ in the point $a \in \CC$ and
$k^{(n)}(g)$ is a simple pole of the integrand lying in the last octant
of the complex $k$-plane for $n\in \NN_+$(see fig.\ref{figura1}), to be evaluated 
in the next section.  

In general, the contribution $\psi^{(l)}_{ex}(x,t;g)$ exhibits an exponential decay, 
while the contribution $\psi^{(l)}_{pw}(x,t;g)$ exhibits a power decay as $t \gg 1$. 
Let us consider the above contributions in turn.

\subsubsection{Exponential Contributions}

The explicit expression of the exponential part of the unstable wavefunction 
at time $t\ge 0$ reads:
\beq
\psi^{(l)}_{ex}(x,t;g) = - 2 \pi i\left(\frac{2}{\pi}\right)^{3/2}\sum_{n=1}^{\infty} {\rm Res} 
\left[  (-1)^l l \frac{\sin (k \pi)}{ k^2 - l^2 } 
\, \frac{ \sin(k x) \, \exp(- i k^2 t )}{ 4 a(k,g) b(k,g) }
; \, k^{(n)}(g) \right] .
\eeq
The Hamiltonian is hermitian (physical case) for real $g$ only, 
which we assume from now on.
The integrand (the first argument in the square bracket above), 
as a function of the complex $k$ variable, 
has removable singularities at the positive integers, $k=l$, and pole singularities
corresponding to the zeroes of the functions $a(k,g)$ and $b(k,g)$
constrained by conditions:
\beq\label{poles2}
\mathrm{Im} \, k^{(n)}(g) \, < \, 0  ; \, \, \, \, \, \, \, \, \, \, \,   
\mathrm{Re} \, k^{(n)}(g) \, > \, |\mathrm{Im} \, k^{(n)}(g)| \, .
\eeq
The transcendental equation 
\beq
\label{bequation}
b(k,g) = 0
\eeq
has simple zeroes for $|g| \ll 1$ of the form
\beq
\label{insert}
k^{(n)}(g) = n - n g + n g^2 - i \pi  n^2 g^2 + 
\mathcal{O}\left( g^3  \right) ,
\eeq
where $n$ is a nonzero integer. All these zeroes lie in the lower half
of the complex $k$-plane, i.e. have $\mathrm{Im} \, k^{(n)}(g) < 0$,
and satisfy also the second condition in eq.(\ref{poles2}) for $n>0$.
In general, the function $k^{(n)}(g)$ is the branch with $k^{(n)}(0)=n$ 
of the multi-valued analytic function $k(g)$ satisfying $b(k(g),g) = 0$.
Numerical computation actually shows that conditions (\ref{poles2}) remain 
satisfied up to values of $g$ of order one.
The zeroes leave the last octant ($-\pi/4 < \theta < 0$) for very large values of $|g|$, 
where the unstable-state description becomes irrelevant. 
Because of eq.(\ref{sim2}), which for real $g$ reads
\beq
a(k,g) = b(k^*,g)^* ,
\eeq
the zeroes of the equation $a(k,g)=0$
are complex conjugates of the ones of eq.(\ref{bequation}), 
therefore lie in the upper half $k$-plane and consequently do 
not enter the residue sum.

\noindent
The only non-trivial residue to evaluate is therefore:
\bea
{\rm Res}\left[\frac{1}{b(k,g)} ; \, k^{(n)}(g) \right]
&=& \lim_{ k \to k^{(n)}(g) } 
\frac{ k - k^{(n)}(g) } { b(k,g) } 
\, = \, 
\frac{1}{\left. (\partial b/ \partial k)(k,g) \right|_{k=k^{(n)}(g)}  } \, 
\nonumber\\
&=& \frac{-2 i \, g \, k^{(n)}(g)}{1 + g \big[ 1 - 2\pi i \, k^{(n)}(g) \big] } \, , 
\eea
where, after the evaluation of the derivative, we have simply replaced 
$k \to k^{(n)}(g)$ and used the relation
\beq
\label{expsolve}
\exp \big[ 2 \pi i k^{(n)}(g)  \big] \, =  \, 1 - 2 \pi i g k^{(n)}(g) \, ,
\eeq
which is true for any solution of eq.(\ref{bequation}).
We then have the following exact expression in terms of the zero set $\{k^{(n)}(g)\}$:
\bea
\psi^{(l)}_{ex}(x,t;g) 
 &=& 
- 2 \pi i\left(\frac{2}{\pi}\right)^{3/2} (-1)^l l \sum_{n=1}^{\infty} 
   \frac{1}{ 4 a\big[ k^{(n)}(g),g \big] (\partial b/\partial k)\big[ k^{(n)}(g), g \big] } 
\, \frac{\sin \big[ k^{(n)}(g) \pi \big] }{k^{(n)}(g)^2-l^2} \times
\nonumber\\
&&~~~~~~~~~~~~~~~~~~~~~~~~~~~~~~~~~~~~~~~~~
\times \sin \big[ k^{(n)}(g) x \big] \, E^{(n)}(t;g)
\nonumber\\
\label{expfinale}
&=&
\sqrt{\frac{2}{\pi}} \, (-1)^l 2 l g 
\sum_{n=1}^{\infty} (-1)^n    
\frac{ k^{(n)}(g) \, \big[ 1 - 2 \pi i \, g \, k^{(n)}(g) \big]^{1/2} }
{\big[l^2 - k^{(n)}(g)^2\big] \big\{ 1 + \big[ 1 - 2 \pi i \, k^{(n)}(g) \big] g \big\}  } \times
\nonumber\\
&& ~~~~~~~~~~~~~~~~~~~~~~~~~~~~~~~~~ \, \, \, \, \, 
\times \sin\left[ k^{(n)}(g) \, x \right] \, E^{(n)}(t;g) \, ,
\eea
where we have defined the time evolution factors 
\beq 
\label{timefact}
E^{(n)}(t;g) \, \equiv \, 
\exp\left[ - i \, \varepsilon^{(n)}(g) \, t \right] \, = \, 
\exp\left[ - i \, \omega^{(n)}(g) \, t - \, \frac{1}{2} \Gamma^{(n)}(g) t \right] \, .
\eeq
Since the energies are complex for $g\ne 0$, on the last member we have split them into 
real and imaginary parts as:
\beq
\varepsilon^{(n)}(g) =  \left( k^{(n)}(g) \right)^2 =  \omega^{(n)}(g) \, - \, \frac{i}{2} \Gamma^{(n)}(g) ,
\eeq
where $\omega^{(n)}(g)$ is the frequency and $\Gamma^{(n)}(g)$ is the decay
width of the pole state $n$. Note that $E^{(n)}(0;g)=1$, as it should.
In eq.(\ref{expfinale}) we have chosen the principal branch of the complex square root,
$-\pi < {\mathrm arg } z \le \pi$ ($1^{1/2}=1$).
In deriving the last member in  eq.(\ref{expfinale}) we have also used a relation
obtained by taking the square root of eq.(\ref{expsolve}):
\beq
\exp\big[ i \pi k^{(n)}(g) \big] \, =  \, 
(-1)^n  \big[ 1 - 2 \pi i \, g \, k^{(n)}(g) \, \big]^{1/2} \, .
\eeq
The sign in front of the square root is fixed by taking the limit 
$g\to 0$ on both sides, i.e. by setting $g=0$ and replacing 
$k^{(n)}(g) \to  n$.
The equality between the first and the last member in eq.(\ref{expfinale}) 
can be written in compact form as:
\beq
\label{expfinalefinale}
\psi^{(l)}_{ex}(x,t;g) = 
\sqrt{\frac{2}{\pi}} \, (-1)^l 2 l g 
\sum_{n=1}^{\infty} (-1)^n F_l\left[ k^{(n)}(g); g \right]
\, \sin\left[ k^{(n)}(g) \, x \right] \, E^{(n)}(t;g) ,
\eeq
where
\beq   
F_l[z;w] \, \equiv \, \frac{ z \left( 1 - 2 \pi i z w \right)^{1/2} }{ 
\left(l^2 - z^2 \right) \left[ 1 + \left( 1 - 2 \pi i \, z \right) w \right]  } .
\eeq
Once the poles $\{ k^{(n)}(g) \}$ have been exactly evaluated (with
numerical methods for example) for a large set of integers, 
$ 1 \le n \le N$ with $N \gg 1$,
one can insert them in the known function $F_l$ on the r.h.s. of 
eq.(\ref{expfinalefinale}). This way one obtains an (almost) exact 
evaluation of the exponential part of the wavefunction.
In the next sections, however, we present an expansion for $g\ll 1$
which allows for explicit analytic expressions.
Equation (\ref{expfinale}) is conveniently rewritten as:
\beq
\label{defVtheta}
\psi^{(l)}_{ex}(x,t;g) =  
\sum_{n=1}^{\infty} V(g)_{l n} \theta^{(n)}(x,t;g) \, ,
\eeq
where the entries of the mixing matrix $V(g)$ read
\beq
\label{defV1}
 V(g)_{l n} \equiv  \,       
\frac{ g (-1)^{l+n} 2 l k^{(n)}(g) \, \big[ 1 - 2 \pi i \, g \, k^{(n)}(g) \big]^{1/2} }
{\big[l^2 - k^{(n)}(g)^2\big] \big\{ 1 + \big[ 1 - 2 \pi i \, k^{(n)}(g) \big] g \big\}  } \, .
\eeq
We have defined the pole wavefunctions (which evolve diagonally with time):
\beq 
\theta^{(n)}(x,t;g) \, \equiv \, \sqrt{ \frac{2}{\pi} } \, \sin \left[ k^{(n)}(g) x \right] \, E^{(n)}(t;g) \, .
\eeq
For $|g|\ll 1$ there is a similarity between the pole wavefunctions above
and the eigenfunctions in eq.(\ref{diviso}) for $k \simeq n-gn \in \RR$.
Let us stress however the differences between the ``true'', exact eigenstates,
having real energies and lying in the continuum spectrum,
and the resonance states, normalizable states with complex energy 
describing dynamics in a simple but approximate way for a finite amount 
of time only \cite{anderson}.

\subsubsection{Matrix Notation}

To simplify formulae, it is convenient to introduce matrix notation.
Let us define an infinite column vector containing all the pole states
\beq
\label{labtheta}
\Theta(x,t;g) \, \equiv \, \left(
\begin{array}{c}
\theta^{(1)}(x,t;g)
\\
\theta^{(2)}(x,t;g)
\\
\cdots
\\
\theta^{(n)}(x,t;g)
\\
\cdots
\end{array}
\right)
\eeq
and an infinite diagonal matrix representing the evolution of the pole states
\beq
{\cal E}(t;g) \, \equiv \, {\rm diag}
\left[ E^{(1)}(t;g), E^{(2)}(t;g), \cdots , E^{(n)}(t;g), \cdots \right] \, . 
\eeq
In more standard notation:
\beq
{\cal E}(t;g) \, = \, \left(
\begin{array}{cccc}
E^{(1)}(t;g) & 0  & \cdots & 0 
\\
0 & E^{(2)}(t;g) & \cdots & 0
\\
\cdots & \cdots & \cdots & \cdots
\\
\cdots & \cdots & E^{(n)}(t;g) & \cdots
\\
0 & 0 & \cdots & \cdots
\end{array}
\right) \, .
\eeq
The temporal evolution of the pole states can be rewritten in matrix notation as:
\beq
\label{polevol}
\Theta(x,t;g) \, = \, {\cal E}(t;g) \, \Theta(x,0;g) \, .
\eeq
Similarly, let us define an infinite column vector containing the metastable 
wavefunctions, 
\beq
\label{vecPsi}
\Psi(x,t;g) \, = \, \left(
\begin{array}{c}
\psi^{(1)}(x,t;g)
\\
\psi^{(2)}(x,t;g)
\\
\cdots
\\
\psi^{(n)}(x,t;g)
\\
\cdots
\end{array}
\right) \, ,
\eeq
as well as the vectors $\Psi_{ex}(x,t;g)$ and $\Psi_{pw}(x,t;g)$ containing the
exponential and power parts respectively\footnote{
In general, let us denote the vectors with uppercase Greek letters and their components
with the corresponding lowercase letters.}. 
Eq.(\ref{residue}) now reads:
\beq
\Psi(x,t;g) \, = \, \Psi_{ex}(x,t;g) + \Psi_{pw}(x,t;g) \, .
\eeq
Setting $t=0$ in eq.(\ref{vecPsi}), one obtains a list of the initial conditions 
for all $l=1,2,3,\cdots$ (see eq.~(\ref{initial})):
\beq
\label{initialvec}
\Psi(x,0) \, = \, \sqrt{\frac{2}{\pi}} \theta(\pi-x) 
\left(
\begin{array}{c}
\sin(x)
\\
\sin(2 x)
\\
\cdots
\\
\sin(n x)
\\
\cdots
\end{array}
\right) \, .
\eeq
Eq.~(\ref{defVtheta}) reads in new notation:
\beq
\label{linear}
\Psi_{ex}(x, t ; g ) \, = \, V(g) \,\, \Theta(x,t;g) \, = \, V(g) \,\, {\cal E}(t;g) \,\, \Theta(x,0;g) . 
\eeq
Let us remark that since there are no power corrections in time, eq.~(\ref{linear}) does not reproduce 
the initial value in eq.~(\ref{initialvec}) for $t=0$.

\begin{figure}[ht]
\begin{center}
\includegraphics[width=0.5\textwidth]{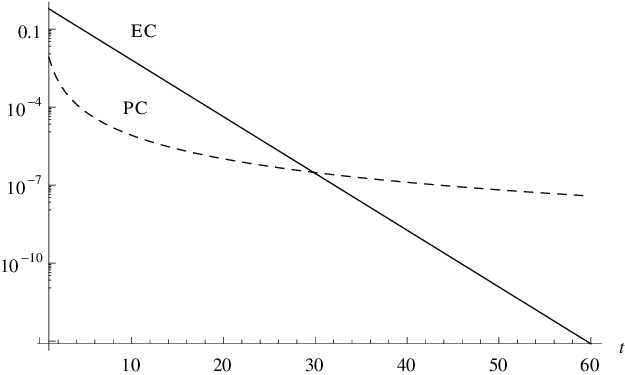}
\footnotesize\caption{
\label{figura2}
\it Time evolution of the modulus square integrated over the cavity
of the exponential contribution (EC, continuous line), 
$\int_0^\pi | \psi^{(1)}_{ex}(x,t;g) |^2 dx$,
and power contribution (PC, dashed line),
$\int_0^\pi | \psi^{(1)}_{pw}(x,t;g) |^2 dx$,
to the wavefunction of the fundamental state 
$\psi^{(1)}(x,t;g)$ for $g=0.2$ (see eq.(\ref{residue})). 
Up to $t \approx 30$ the exponential contribution dominates over the
power one, which controls the asymptotic behavior. 
The scale on the vertical axis is logarithmic.}
\end{center}
\end{figure}

\subsection{Power Contributions}

The integral $\psi^{(l)}_{pw}(x,t;g)$, over the ray ($0,\, \infty \, e^{-i\pi/4}$)
in complex $k$-plane,
can be exactly evaluated with numerical methods without problems for any
time $t \ge 0$, as it does not involve any oscillation. It is also convergent at the
initial time $t=0$; in other words, the decomposition in eq.(\ref{residue})
does not spoil the convergence at $t=0$.
However, for large times, $t \gg 1$, the integral takes the dominant contribution 
from a neighborhood of $k = 0$, 
where the integrand is analytic and can therefore be expanded in powers of $k$:
\beq
\label{pj}
p^{(l)}(k;x,g) \, = \, \frac{g^2}{(1+g)^2}\sum\limits_{j=1}^{\infty}p^{(l)}_j(x,g) \, k^{2j} \, .
\eeq
The first two coefficients explicitly read:
\bea
p_1^{(l)}(x,g) &=& \frac{ (-1)^{l+1} }{l} \, \pi x \, ;
\\
p_2^{(l)}(x,g) &=& \frac{ (-1)^{l+1} }{l} \pi x
\left[
\frac{1}{l^2} + \frac{\pi^2}{6} + \frac{2}{3} \frac{\pi^2 g}{1+g} 
- \frac{\pi^2 g^2}{(1+g)^2} - \frac{x^2}{6}
\right] \, .
\eea
Replacing the series on the r.h.s. of eq.(\ref{pj}) into the integral 
over $k$ on the r.h.s. of eq.(\ref{powercorr}), exchanging the integral
with the series and performing the change of variable $\nu = k^2 \, t$, 
one obtains the following asymptotic expansion:
\bea
\psi^{(l)}_{pw}(x,t;g)
& \approx & \frac{\sqrt{2}}{\pi^{3/2}}\frac{e^{-i\pi/4}g^2}{(1+g)^2}
\sum\limits_{j=1}^{\infty}\frac{(-i)^j \, p_j^{(l)}(x,g)}{t^{j+\frac{1}{2}}}
\int\limits_{0}^{\infty} d\nu \, \nu^{j-\frac{1}{2}} \, e^{-\nu}
\\
&=&
\frac{\sqrt{2}}{\pi}\frac{e^{-i\pi/4}g^2}{(1+g)^2}\sum\limits_{j=1}^{\infty}\frac{(-i)^j(2j-1)!!}{2^j}
\frac{p_j^{(l)}(x,g)}{t^{j+\frac{1}{2}}} , ~~ \, \, \, 0 \le x \le \pi , \, \, ~ t \gg 1 ,
\eea
whose first two terms read:
\bea
\psi^{(l)}_{pw}(x,t;g) &\approx&
\frac{ e^{i\pi/4} }{\sqrt{2}} \frac{ (-1)^l }{l} \frac{g^2}{(1+g)^2} \frac{x}{ t^{3/2} } \times
\nonumber\\
&\times& \Bigg\{ 
1 \, - \,  
\frac{3 i}{2t} 
\Bigg[
\frac{1}{l^2} + \frac{\pi^2}{6} + \frac{2}{3} \frac{\pi^2 g}{1+g}
- \frac{\pi^2 g^2}{(1+g)^2} - \, \frac{x^2}{6}
\Bigg]
+ \mathcal{O} \left( \frac{1}{t^2} \right)
\Bigg\} \, . 
\eea
Let us make a few remarks.
The physical interpretation of the small $k$ expansion
is that states with very low momenta are produced
in the decay at asymptotic times \cite{winter}.
The above asymptotic expansion is uniformly valid for all $g \ge 0$, since 
the coefficients $p_j^{(l)}(x,g)$ are uniformly bounded in that region 
(see eq.~(\ref{pj})).
The exponent $3/2$ controlling the power decay, $\psi_{pw} \approx 1/t^{3/2}$,
does not depend on $l$ and $g$; power corrections however vanish for
$g \to 0$ (impermeable cavity).

\noindent
We are in complete agreement with \cite{winter} as far
as the asymptotic power behavior in time is concerned;
however we remark that our results for the power corrections in $t$
are valid for any $g$, i.e. do not involve any expansion in $g$.
In particular, one can take the limit $g\to \infty$, in which the
potential barrier disappears. 

\section{Resonances}

For $g\ll 1$, i.e. for weak coupling, there is a large time slice between a 
pre-exponential small-$t$ region \cite{chap4,winter,primo} 
and a post-exponential one related to the power-like decay just discussed,
\beq
\label{diseg}
1 \ll t \lsim \frac{\log(1/g)}{g^2} ,
\eeq
in which the unstable wavefunctions $\psi^{(l)}(x,t;g)$ evolve 
to a good approximation as a superposition of pole states, i.e. of resonances (see fig.\ref{figura2}).
Relation (\ref{diseg}) is a consequence of the first-order results in the
next section.
It is clear that non-exponential contributions do not have a resonance interpretation:
they constitute an intrinsic limit of the scheme.

\subsection{First-order computation, $\mathcal{O}(g)$}

By expanding in powers of $g$ the mixing matrix
\beq
V(g) = \sum_{k=0}^\infty g^k V^{(k)} ,
\eeq
one obtains up to first order:
\bea
\label{V0}
V^{(0)} &=& Id ;
\\ 
\label{V1}
V^{(1)} &=& - \frac{1}{2} Id + A ,
\eea
where $A$ is the real antisymmetric matrix with entries
\beq
A_{l,n} \equiv (-1)^{l+n} \frac{  2 \, l \, n }{ l^2 - n^2 } \, \, \, \, {\rm for} \, \, \, \, \, l \ne n
\eeq
and $A_{l,l}=0$.
The frequencies and widths entering the pole wavefunctions have the following lowest-order
expressions:
\bea
\omega^{(n)}(g) &\equiv& + \, \left[ \mathrm{Re} \, k^{(n)}(g) \right]^2 \, - \, 
\left[ \mathrm{Im} \, k^{(n)}(g) \right]^2 
\nonumber\\
&=& n^2 \left( 1 - 2g \right) \, + \, \mathcal{O}\left( g^2 \right) \, ;
\\
\Gamma^{(n)}(g) &\equiv& - \, 4 \, { \mathrm{Re} \, k^{(n)}(g) } \, { \mathrm{Im} \, k^{(n)}(g) }
\nonumber\\
\label{width2e3}
&=&  4 \pi n^3 g^2 \, + \, {\mathcal O}\left( g^3 \right) \, .
\eea

\noindent
Since it is convenient to have some freedom in the normalization of the pole 
states, let us introduce the renormalization constants
\beq
Z^{(n)}(g) = 1 + \sum_{k=1}^\infty g^k \, z^{(n)}_k ,
\eeq
where $z^{(n)}_k$ are (in general complex) coefficients, which depend in general on $n$: they have to be 
determined by imposing chosen renormalization conditions.
We define the renormalized pole states $\xi^{(n)}(x,t;g)$ 
by dividing $\theta^{(n)}(x,t;g)$ by $Z^{(n)}(g)$:
\beq
\xi^{(n)}(x,t;g) \, \equiv \, \frac{ \theta^{(n)}(x,t;g) }{ Z^{(n)}(g) } 
= \frac{ 1 }{ Z^{(n)}(g) }  \, \sqrt{ \frac{2}{\pi} } \, \sin \left[ k^{(n)}(g) x \right] \, E^{(n)}(t;g) \, 
\eeq
and define the renormalized mixing matrix entries $U(g)_{l,n}$ by multiplying $V_{l,n}(g)$ 
by the same factor,
\beq
U(g)_{l,n} =  V(g)_{l,n} \, Z^{(n)}(g)
\eeq
(no sum over $n$ is implied).
In order to introduce a matrix notation, let us represent the renormalization constants 
through the diagonal matrix
\beq
\label{defZcal}
{\cal Z}(g) \equiv { \rm diag } \left[ Z^{(1)}(g), Z^{(2)}(g), \cdots ,  Z^{(n)}(g), \cdots \right] \, .
\eeq
In components, eq.(\ref{defZcal}) reads:
\beq
{\cal Z}(g)_{l,n} = \delta_{l n}  Z^{(n)}(g) ,
\eeq
where $\delta_{l n} = 1$ for $l = n$ and zero otherwise is the Kronecker $\delta$.
The matrix renormalization constant also possesses a power-series expansion: 
\beq
{\cal Z}(g) = Id + \sum_{k=1}^\infty g^k {\cal Z}^{(k)} ,
\eeq
where ${\cal Z}^{(k)}$ are diagonal matrices.
The renormalized mixing matrix then reads
\beq
U(g) \, = \, V(g) \, {\cal Z}(g) .
\eeq
In terms of the renormalized pole states, eq.(\ref{linear}) reads:
\beq
\label{linear2}
\Psi_{ex}(x, t ; g ) \, = \, U(g) \,\, \Xi(x,t;g) \, = \, U(g) \,  {\cal E}(t;g)   \, \Xi(x,0;g) .
\eeq
By multiplying with each other the power series's of $V(g)$ and ${\cal Z}(g)$, 
one obtains the power expansion for $U(g)$,
\beq
U(g) \, = \, \sum_{k=0}^\infty g^k U^{(k)} ,
\eeq
where
\beq
U^{(n)} \, = \, \sum_{k=0}^n V^{(k)} {\cal Z}^{(n-k)} .
\eeq
Because of the change of wavelength due to the interaction, i.e. to $g\ne 0$,
the pole states $\theta^{(n)}(x,t;g)$ are not normalized to one at the initial time
$t=0$, unlike the initial conditions. If we introduce normalized pole states,
i.e. satisfying the condition
\beq
\label{normto1}
\int_0^\pi | \xi^{(n)}(x,0;g) |^2 dx \, = \, 1 \, ,
\eeq
we obtain
\beq
Z^{(n)}(g) \, = \, 1 + \frac{g}{2} \, + \mathrm{O}\left( g^2 \right) \, .
\eeq
Therefore, up to first order:
\bea
U^{(0)} &=& Id ;
\\
U^{(1)} &=& A \, .
\eea
It is remarkable that condition (\ref{normto1}) has the effect of removing the diagonal
contributions from the first-order mixing matrix $U^{(1)}$.
Let us also note that the matrix $U(g) = Id + g A + \cdots$ represents
an infinitesimal rotation in the infinite dimensional vector space 
of the normalized pole states. So the natural question, to be treated in
the next section, is:"what happens in higher orders in $g$?''.

\subsubsection{Comparison with Winter results}

\begin{figure}[ht]
\begin{center}
\includegraphics[width=0.5\textwidth]{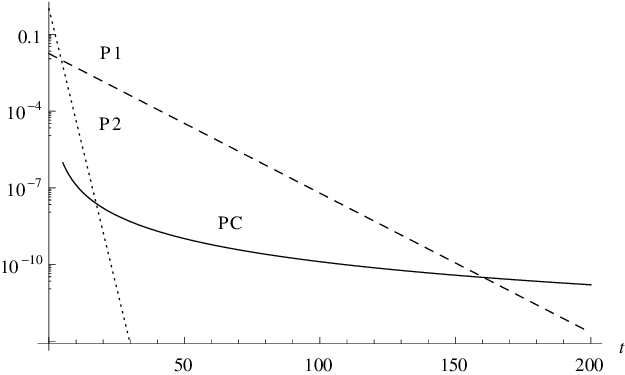}
\footnotesize
\caption{
\label{figura3}
\it Time evolution of the contributions to the $l=2$, 
i.e. first excited, state
for $g=0.1$.
Dotted line: second pole contribution (P2);
Dashed line: first pole contribution (P1);
Continuous line: power contribution (PC).}
\end{center}
\end{figure}

We are in disagreement with \cite{winter} regarding the exponential behavior of the excited
metastable states, i.e. of $\psi_{ex}^{(l)}(x,t;g)$ with $l>1$. Let us show that in detail.
Eq.(\ref{linear2}) reads in components:
\bea
\label{nuova}
&&\psi_{ex}^{(l)}(x,t;g) = 
\xi^{(l)}(x,0;g) \exp\left[ - i \, \omega^{(l)}(g) \, t - \, \frac{1}{2} \Gamma^{(l)}(g) t \right] \, +
\nonumber\\
&& \, 
+ \sum_{n \, \ne \, l}^{1,\, \infty} \, g \frac{  (-1)^{l+n} 2 \, l \, n }{ l^2 - n^2 } \, 
\xi^{(n)}(x,0;g) \exp\left[ - i \, \omega^{(n)}(g) \, t - \, \frac{1}{2} \Gamma^{(n)}(g) t \right] \, .
\eea
As usual, $l$ labels the initial state and $n$ the pole state.
The diagonal term on the r.h.s. of the above equation 
(the one with $n=l$) is in agreement with the r.h.s. of eq.(2a) in \cite{winter}. 
In \cite{winter} however, the non-diagonal contributions (the ones with $n\ne l$),
entering the sum on the r.h.s., are not included. These terms have a 
coefficient suppressed by a power of $g \ll 1$ compared to the
diagonal one, but have a slower exponential decay for
$n<l$ ($\Gamma^{(n)} \propto n^3$, see eq.(\ref{width2e3})), 
and therefore dominate in the exponential time region (i.e. before
power-effects take over).
A reasonable approximation is to truncate the sum on the r.h.s. of eq.(\ref{nuova})
to $n<l$, as contributions with $n>l$ are suppressed both by a power of $g\ll 1$ 
and by the large widths.
For example, for $l=2$ eq.(\ref{nuova}) may be approximated 
by neglecting higher poles as
\bea
&&\psi_{ex}^{(2)}(x,t;g) \simeq 
\xi^{(2)}(x,0;g) \exp\left[ - i \, \omega^{(2)}(g) \, t - \, 16 \pi g^2 t \right] \, +
\nonumber\\
&&  \, \, \, \, \, \, \, \, \, \, \, \, \, \, \, \, \, \, \, \, \, \,  
- \frac{ 4 }{ 3 } g \, 
\xi^{(1)}(x,0;g) \exp\left[ - i \, \omega^{(1)}(g) \, t - \, 2 \pi g^2 t \right] \, .
\eea
As a measure of the size of the above terms, let us take the square of
the modulus integrated over the cavity ($0<x<\pi$):
\beq
\int_0^\pi | \cdots |^2 dx \, .
\eeq
As shown in fig.\ref{figura3}, for $g=0.1$ there is a large temporal region, 
from $t \simeq 5$ up to $t \simeq 160$,
where the non-diagonal contribution from the first pole,
\beq
\frac{16}{9} g^2 \int_0^{\pi}|\xi^{(1)}(x,0;g)|^2 dx \,    
\exp\left( \, - \, 4 \pi g^2 t \right) \, ,
\eeq
dominates over the diagonal one from the second pole, 
\beq
\int_0^{\pi}|\xi^{(2)}(x,0;g)|^2dx \, \exp\left( \, - \, 32 \pi g^2 t \right) \, ,
\eeq
in the temporal evolution of the first excited state $(l=2)$.
For very large times, $t \gsim 160$, the power contribution dominates over the 
exponential ones and we enter the (true) asymptotic region.
For general $g$, the non-diagonal $l=1$ contribution dominates over the diagonal one
$l=2$ for
\beq
t > t^* \simeq \frac{1}{28 \pi g^2} \log \left( \frac{9}{16 g^2} \right) .
\eeq
For $t=t^*$ the signal-to-noise ratio (i.e. the modulus squared of the wavefunction 
normalized at $t=0$ integrated over the cavity) is decreased from one down to
$\simeq 2 \left( 16 g^2 / 9 \right)^{8/7}$ (the factor two comes from the two
resonances).

To summarize, neglecting the non-diagonal contributions is a reasonable approximation 
only for the time-evolution of the lowest-lying state $\psi^{(1)}(x,t;g)$. 
The presence of the non-diagonal terms shows that the evolution of general unstable
states is far more complicated than as implied by the analysis in \cite{winter}.
As far as we know, the occurrence and the relevance of such off-diagonal terms
has been originally noted in \cite {primo}.
A physical interpretation of such effect will be presented in the next section.

\subsubsection{Physical Interpretation of Pole State Mixing}
\label{fisica}

In order to express the metastable wavefunctions of Winter model 
$\psi^{(l)}(x,t;g)$ in terms of the eigenfunctions of the particle in a box, 
one has to diagonalize the time evolution. That is achieved by ``counter-rotating'' 
the vector containing the initial conditions, i.e. by considering the evolution not of
$\Psi(x,0;g)$ but of
\beq
\label{rotate}
\Phi(x,0;g) \, \equiv \, U^{-1}(g) \, \Psi(x,0;g) \, .
\eeq
By using the first equality in eq.~(\ref{linear2}), it is immediate to show that
\beq
\Phi(x, t ;g) \, =  \, \Xi(x,t;g) = {\cal E}(t;g) \, \Xi(x,0;g) \, .
\eeq
By looking at the vector equation (\ref{rotate}) component by component,
the new initial conditions read:
\beq
\label{rotatebis}
\phi^{(l)}(x,0;g) = \sqrt{\frac{2}{\pi}} \theta(\pi-x) \sum_{n=1}^\infty \big(U^{-1}(g)\big)_{l n}  
 \sin(n x) ,
\eeq
each evolving as a single pole wavefunction:
\beq
\phi^{(l)}(x, t ; g) = \xi^{(l)}(x,t;g) .
\eeq
The ``experimental meaning'' of eq.~(\ref{rotate}) or eq.~(\ref{rotatebis}) is clear:
in order to observe a diagonal time evolution as in the free case (\ref{freeeg}), 
one has to prepare the initial state as the coherent superposition of free
eigenfunctions given by eq.~(\ref{rotate}) or by eq.~(\ref{rotatebis}).
In the case of excited states, $l>1$, the superposition in 
eq.~(\ref{rotate}) or (\ref{rotatebis}) has also the effect of subtracting the
contributions from smaller $l$'s, which decay slower in time
and therefore tend to dominate the evolution, as discussed in \cite{primo}.
If the matrix $U(g)^{-1}$ is computed in an approximate way (as a truncated power 
series in $g$ for example), there is a small residual contamination 
in the time evolution of the $l$-th state from the lower ones,
which becomes substantial asymptotically in time.
In other words, the problem of isolating the $l$-th mode for all times
can in principle be solved only with an exact computation of $U(g)$.

Let us now explicitly evaluate the initial wavefunction which
evolves diagonally in $t$, according to eq.~(\ref{rotatebis}). To order $g$:
\beq
U(g)^{-1} = 1 - g A + \mathcal{O}(g^2) .
\eeq
The sum of the trigonometric series on the r.h.s. of eq.~(\ref{rotatebis})
reads:
\beq
\phi^{(l)}(x,0;g) = \sqrt{\frac{2}{\pi}} 
\left[
\left( 1 - \frac{g}{2} \right) \sin (l x)
- g l x \cos(l x) 
\right] \theta(\pi-x) \, + \, \mathcal{O}\left(g^2\right) .
\eeq
The r.h.s. of the above equation is the expansion to $\mathcal{O}\left(g\right)$ of
\beq
\label{semplice}
\sqrt{\frac{2}{\pi}} \, 
\left( 1 - \frac{g}{2} \right) \, \sin \left[ l ( 1 - g ) \, x \right] 
\theta(\pi-x) \, .
\eeq
It is tempting to think that higher orders in $g$ actually lead to the
result (\ref{semplice}): the term $g^2 A^2/2$ in the second-order correction
$g^2 U^{(2)}$ actually confirms this guess. However, let us warn the reader
that wavefunction renormalization does not account for all the second-order
effects (see next section).

The interpretation of eq.~(\ref{semplice}) is straightforward: the counter-rotation
of the initial wavefunction in index space amounts to the shift of the wave-vector 
$l \to k^{(l)}(g) = l(1-g) + \cdots$ in momentum space, 
with a consequent change of normalization. 
In other words, in order to have a diagonal evolution in $t$ of the initial wavefunction,
the latter has to be {\it prepared}  with the {\it corrected} wave-vector $k^{(l)}(g)$, which is
dynamically generated from $l=k^{(l)}(0)$, the free one. 
Temporal evolution is then simply given by multiplication by the factor $E^{(l)}(t,g)$
in eq.~(\ref{timefact}).

\noindent
Let us note that the wavefunction in eq.~(\ref{semplice}) has a finite jump
$\mathcal{O}\left(g\right)$ at the right border of the cavity, in $x=\pi$.
The Fourier series in eq.~(\ref{rotatebis}) exhibits indeed the Gibbs
phenomenon in $x=\pi$, as the coefficients decay asymptotically $\approx 1/n$
for $n \to \infty$ 
\footnote{The related ``vertical slope''  in $x=\pi$ can be derived by 
differentiating eq.~(\ref{rotatebis}) with respect to $x$ and then 
setting $x=\pi$.}.
It is remarkable that we obtain a discontinuous initial wavefunction, while
the eigenfunctions only have a discontinuous first derivative \cite{winter,flugge,primo}.  
Let us remark however that the results above are obtained by means of
power series in $g$ which we have not shown to be convergent, and are probably
only asymptotic. The Fourier series therefore should be truncated to some
finite order in $g$, regularizing the discontinuity.

\subsection{Second-Order Computation, $\mathcal{O}(g^2)$ }

In this section we push the perturbative expansion for $g \ll 1$ up to 
$\mathcal{O}(g^2)$ included, in order to obtain more accurate results and 
to get some insight into the general structure of the expansion, if any. 
The exact expression in eq.(\ref{defV1}) for the mixing matrix in terms
of the exact solutions $k^{(n)}(g)$ of the equation $b(k,g)=0$ indeed is not
very illuminating.
By inserting the small-$g$ expansion for the poles pushed one order further
with respect to previous section,
\beq
\label{insert2}
k^{(n)}(g) = n - n g + \left( n - i \pi  n^2 \right) g^2 + 
\Big(\frac{4 }{3} \pi ^2 n^3 + 3 i \pi n^2-n\Big) g^3
+ \mathcal{O}\left( g^4  \right) ,
\eeq
we obtain for the mixing matrix:
\bea
V^{(0)} &=& Id ;
\\ 
V^{(1)} &=& A  - \frac{1}{2} Id ;
\\
\label{Vsecond}
V^{(2)} &=&  \frac{1}{2} A^2 - A  + \frac{3}{8} Id + i \pi A H - \frac{3}{2} i \pi H  ,
\eea
where for convenience we have repeated the lowest-order results 
and $H$ is the real diagonal matrix
\beq
H \equiv { \rm diag } \left( 1, 2, 3, \cdots, n , \cdots \right) .
\eeq
The coefficients entering the pole wavefunctions read:
\bea
\omega^{(n)}(g) &=& n^2 \left( 1 - 2g + 3 g^2 \right) \, + \mathcal{O}\left( g^3 \right) \, ;
\\
\label{Gammas}
\Gamma^{(n)}(g) &=&  4 \pi n^3 g^2 
\left(1 \, - \, 4 g \right)\, + \, {\mathcal O}\left( g^4 \right) \, .
\eea

\noindent
Eq.(\ref{Vsecond}) has been obtained by using the explicit expression
\beq
\label{V2}
V^{(2)}_{l n} =
\delta_{l n} \left( \frac{1}{4} - \frac{\pi^2}{6} l^2 - i \pi \frac{3}{2} l \right) 
+  \left( 1 - \delta_{l n} \right)
\left[
\, \frac{ (-1)^{l+n} \, 2 l n }{ l^2 - n^2 } \big( i \pi n  - 1 \big)
+ \, \frac{ (-1)^{l+n+1} \, 2 l n \, (l^2+n^2) }{ (l^2 - n^2)^2 } 
\right] \, , 
\eeq
together with the formula
\beq
\frac{1}{2} \left( A^2 \right) _{l n} = 
\left( 1- \delta_{l n} \right) (-1)^{l+n+1} \, \frac{2 l n \, (l^2+n^2)}{(l^2-n^2)^2}
- \delta_{l n} \left( \frac{\pi^2}{6} l^2 + \frac{1}{8} \right) .
\eeq 
Note that the squared matrix is symmetric, as it should, being the square 
of an antisymmetric matrix.
The last equation has been derived by means of the identities  
\footnote{
These identities are obtained from
$\sum_{k \ne m }^{-\infty,+\infty} 1/(k-m)=0$ and
$\sum_{k \ne m }^{-\infty,+\infty} 1/(k-m)^2=\pi^2/3$
respectively, by splitting the sums into positive and negative 
indices's and rearranging them in order to have a single sum
(the first identity can also be found in \cite{Grad}).
}
\beq
\sum_{k\ne m}^{1,\infty} \frac{1}{k^2-m^2} = \frac{3}{4m^2}
~~~~~{\rm and}~~~~~
\sum_{k\ne m}^{1,\infty} \frac{k^2}{\left(k^2-m^2\right)^2} = 
\frac{\pi^2}{12} + \frac{1}{16 m^2} ,
\eeq
which hold for $m$ a positive integer.
Let us remark that it is not trivial that the matrix $A^2$ does exist,
as its entries involve the summation of infinite series, which
in effect turn out to be (absolutely) convergent.
By looking at the asymptotic form of the coefficients of $A$ and $A^2$
given above, it is straightforward to show that $A^3$ and $A^4$ also
exist. We expect that all the positive powers of $A$ do exist. 

Let us now discuss renormalization at second order.
The explicit expansion of $U(g)$ up to ${\mathcal O}\left(g^2\right)$ reads:
\bea
U^{(0)} &=& Id ;
\\
U^{(1)} &=& V^{(1)} + {\cal Z}^{(1)} 
\nonumber\\
&=& A - \frac{1}{2} Id + {\cal Z}^{(1)} ;
\\
U^{(2)} &=& V^{(2)} + V^{(1)} {\cal Z}^{(1)} + {\cal Z}^{(2)} 
\nonumber\\
&=& \frac{1}{2} A^2 - A  + \frac{3}{8} Id + i \pi A H - \frac{3}{2} i \pi H
+ \left (A - \frac{1}{2} Id \right) {\cal Z}^{(1)}  
+ {\cal Z}^{(2)} \, .
\eea
As shown in previous section, if we set
\beq
{\cal Z}^{(1)} = \frac{1}{2} Id ,
\eeq
we obtain at first order
\beq
U^{(1)} = A ,
\eeq
so that at second order we get:
\beq
U^{(2)} = \frac{1}{2} A^2 - \frac{1}{2} A  + \frac{1}{8} Id - \frac{3}{2} i \pi H + i \pi A H + {\cal Z}^{(2)} .
\eeq
We may set for example
\beq
{\cal Z}^{(2)}  =  - \frac{1}{8} Id + \frac{3}{2} i \pi H ,
\eeq
to give
\beq
U^{(2)} = \frac{1}{2} A^2 - \frac{1}{2} A  + i \pi A H .
\eeq
The renormalized mixing matrix then finally reads:
\beq
\label{finalU}
U(g) = Id + g A + \frac{1}{2} g^2 A^2 - \frac{1}{2} g^2 A  + i \pi g^2 A H +  {\mathcal O}\left(g^3\right) .
\eeq
Let us notice that, by introducing the renormalized coupling
\beq
g_r \equiv g - \frac{1}{2} g^2  +  {\mathcal O}\left(g^3\right) ,
\eeq
the formula above can be simplified a bit:
\beq
\label{finalU2}
U(g_r) = Id + g_r A + \frac{1}{2} g_r^2 A^2 + i \pi g_r^2 A H +  {\mathcal O}\left(g_r^3\right) .
\eeq
The inverse matrix is given up to second order in $g_r$ by:
\beq
\label{finalU^-1}
U^{-1}(g_r) = Id - g_r A + \frac{1}{2} g_r^2 A^2 - i \pi g_r^2 A H 
+ {\mathcal O}\left(g_r^3\right) .
\eeq
As anticipated in previous section, while the term  $1/2 \, g_r^2 A^2$
can be absorbed in the box eigenfunctions by means of wave-vector renormalization, 
that is not true for the term $- i \pi g_r^2 A H$. The latter has a large size,
\beq
\left( A H \right)_{l,n} = \frac{ (-1)^{l+n} \, 2 l n^2 }{ l^2 - n^2 } \, \, \, 
{\rm for } \, l \ne n \, \, \, {\rm and} \, \, 0 \, \, {\rm otherwise} \, ,
\eeq
and produces also a highly singular behavior in the counter-rotated 
box eigenfunctions considered in the previous section, $U(g)^{-1}\Psi(x,0;g)$,  
because
\beq
(A H)_{l,n} \, = \, {\mathcal O}(1) \, \, \, {\rm for} \, n\to\infty \, \, (l \, {\rm fixed}).
\eeq
The detailed investigation of such effects, for which we are not able  
to provide at present a physical interpretation, requires the study of the convergence 
properties of the series in $g$ involved, which is beyond the scope of the present paper.

\subsubsection{Exponentiation}

The first three terms on the r.h.s. of eq.(\ref{finalU2}) are actually the expansion of
\beq
\exp[g_r A] \, = \, Id + g_r A + \frac{1}{2}  g_r^2 A^2 + {\mathcal O}\left(g_r^3\right) ,
\eeq
so it is not difficult to conjecture that higher orders in $g_r$ will lead to the exponential 
above. The problem is that we are not sure that the conjectured exponentiation is "legitimate'',
i.e. that it includes {\it all} the {\it leading} terms order by order in $g$. 
To ${\mathcal O}\left(g^2\right)$ we encountered indeed the ``big'' term $A H$, 
which will presumably produce iterates of similar size in higher orders,
which we are unable to control.
Furthermore, since ${\cal Z}^{(2)}$ is a diagonal matrix, whatever value is chosen for it, 
we cannot cancel the term $A H$ in $U^{(2)}$ with an ad-hoc renormalization condition.
A third-order computation in $g$, which is in principle straightforward while technically
cumbersome, could probably reveal further structure of the perturbative expansion.

\section{Discussion}

Let us now discuss the renormalized wavefunctions $\phi^{(l)}(x,t;g)$.
The main qualitative difference between the free case and the interacting one
is that in the latter case there are non-zero widths.
The appearance of an imaginary part in the {\it ab initio} real energy
is a second order effect in $g$\footnote{
Non zero widths are clearly not in contradiction with the unitarity 
of the fundamental theory because we are looking at a subsystem,
an ``open'' system \cite{hatano,chap7}.
}.
Once a non-zero width is allowed, the key point is that the $\phi^{(l)}(x,t;g)$'s
have a similar form to the eigenfunctions of the free system $\psi_0^{(l)}(x,t)$ 
in eq.~(\ref{freeeg}). The differences between the free case 
and the interacting one, as long as $0<g \ll 1$, can be relegated to small modifications 
of the parameters entering the free wavefunctions $\psi_0^{(l)}(x,t)$.
In other words, switching on the interaction, i.e. going from $g = 0$ to $0 < g \ll 1$, 
produces finite renormalizations only.
Let us discuss these renormalizations in turn:
\begin{enumerate}
\item
the normalization coefficient $Z^{(l)}(g)$ has a modulus greater than one
for $0< g \ll 1$ and reduces to $1$ in the free case $g=0$; 
it has a first-order correction in $g$ and is the analog of the 
field renormalization constant $Z$ in quantum field theory \cite{primo}. 
Unlike the most common cases (QED for example), 
$Z^{(l)}(g)$ is not real because the one-particle states are unstable;  
\item
the wave-vector $k^{(n)}(g)$ is renormalized to first order in $g$ by the interaction
and reduces to the free case for $g \to 0$: $k^{(l)}(0)=l$.
It acquires an imaginary part at second order in $g$, related to the decay width.
That implies the disappearance of the node of the wavefunction  
around $x=\pi$ and a (small) exponential growth of $\phi^{(l)}(x,t;g)$
by going from the impermeable wall in $x=0$ toward the permeable one in $x=\pi$;
\item
the real part of the energy $\omega^{(l)}(g)$ is also renormalized to first order in $g$ 
by the interaction and reduces to the free case for $g \to 0$: $\omega^{(l)}(0)=l^2$.
Note that the free dispersion relation $\omega=k^2$ is not renormalized at first order.
\end{enumerate}
Let us make a few remarks. 
\begin{itemize}
\item
We do not expand in powers of $g$ the wavefunctions $\phi^{(l)}(x,t;g)$, 
but only the parameters $k^{(n)}(g),\,\omega^{(n)}(g),$  etc. entering them through the 
functions appearing in $\psi_0^{(l)}(x,t)$.
That implies that we are resumming classes of higher order corrections in $g$ in the
wave function, in the spirit of renormalization in quantum field theory 
\cite{hint,lepage,ms,zinnjustin} and statistical mechanics \cite{ma},
or the method of multiple scales in classical physics \cite{landau1,ingegnere};
\item
The decay widths grow faster with increasing $n$ than the frequencies: 
\beq
\Gamma^{(n)}(g) \propto n^3, ~~~~~ \omega^{(n)}(g) \propto n^2.
\eeq
Since our renormalized theory has meaning only for
\beq
\label{qualcond}
\Gamma^{(n)}(g) \ll \omega^{(n)}(g),
\eeq
we cannot take $n$ too large. Therefore, while in principle the state vectors
and the evolution/mixing matrices are infinite, in practice for any fixed $g$ one has 
to make a truncation in $n$ according to the condition (\ref{qualcond}) 
\footnote{
\label{nfl}
In general, in the theory of elementary excitations, one only considers 
states lying slightly above the ground state \cite{anderson}.
In the case of normal Fermi liquids, for example, one only considers electrons slightly above the 
Fermi surface, i.e. with $\epsilon_k \equiv k^2/(2m) - E_F \ll E_F$,
where $E_F$ is the Fermi energy.
An electron slightly above the Fermi surface is an unstable state because it can hit an 
electron slightly below it creating a pair. In this case $\Gamma_k \propto \epsilon_k^2$,
i.e. the width increases with the square of the excitation energy.
}.
This limitation is also reasonable from physics viewpoint:
high energy particles pass through the barrier in $x = \pi$ without difficulty
and therefore there is no sense in including them to describe the dynamics 
inside the cavity.
By restricting on $n$ one is also cutting off small wavelengths $\lambda \lsim 2\pi/n$
and therefore is limiting space resolution. 
\end{itemize}

\section{Conclusions}

In this work we have shown that the evolution according to Winter
model of the eigenfunction of a particle inside an impermeable cavity
(i.e. a box) with {\it any} quantum number $l=1,2,3,\cdots$ is not controlled 
asymptotically by the corresponding $l$-th resonance, as intuitively expected
and as stated in \cite{winter}, but {\it always} by the {\it first} resonance, 
$l=1$.
This phenomenon originates from $\mathcal{O}(g)$ coupling terms 
between the eigenfunctions of the particle in a box 
("box eigenfunctions'' hereafter) and the resonances, which we have 
evaluated with next-to-leading order accuracy, i.e. up to second order in 
the coupling $g\ll 1$. 
Because of this mixing, metastable dynamics is far more 
complicated than as implied by Winter results and common arguments. 
Roughly speaking, the idea suggested by our results is that the box 
eigenfunctions are not "natural" initial states as far as decay is 
concerned: ``natural'' initial states are dynamically generated
from the box eigenfunctions.
With the exception of the fundamental state, such box eigenfunctions do not even 
approximately evolve as single elementary excitations, as claimed by 
Winter, but as coherent superpositions of many elementary excitations. 
Furthermore, mixing effects are quite large, as mixing matrix entries 
decay rather slowly as we move away from the main diagonal (as $1/n$
to $\mathcal{O}(g)$, where $n$ is the distance from the diagonal, see eqs.(52) and (89)).
The physical picture is that time evolution produces, in addition 
to the expected decay of box eigenfunctions, also a finite "rotation" of 
them, represented by the infinite matrix $U(g)$. 
Therefore, in order to have a "natural'' initial state evolving simply, 
i.e. diagonally, with time, one has to 
"counter-rotate" the box eigenfunctions, by means of the matrix 
$U^{-1}(g)$.
One of the main dynamical effects of such a counter-rotation is to 
adjust the wavelengths of the box eigenfunctions to those of the  
elementary excitations of the model. In physical terms, the time 
evolution of box eigenfunctions produces a rearrangement of their 
wavelengths to the characteristic wavelengths of the system.
Let us stress however that wavelength renormalization does not
exhaust the effects contained in the matrix $U^{-1}(g)$ (a complete 
physical interpretation is still missing).
Since Winter model is a limiting case of many different models, 
it is clear that the occurrence of nondiagonal terms has to
be a general phenomenon in metastable systems. 
The implications of the mixing terms for more complex models
are, as far as we know, still to be investigated and could
modify their current understanding.  
Let us stress that it would have been quite difficult to imagine the 
existence of such mixing terms in more phenomenologically 
relevant models, but also much more complicated, without our exhaustive 
analytic study of Winter model. 

We have also shown that the resonant states of Winter model can be 
related to the box eigenfunctions by means of renormalization of the 
parameters entering the ``free'' eigenfunctions, after allowing for 
non-zero widths.

\end{document}